\documentclass[journal]{IEEEtran}
\usepackage{cite}

\ifCLASSINFOpdf
    \usepackage[pdftex]{graphicx}
    \usepackage{pgfplots}
    \usepackage{grffile}
    \pgfplotsset{compat=1.14}
    \usepackage{lscape}
\usepackage{xcolor}
    \usepackage{array, booktabs, makecell, tablefootnote}
    \usepackage{tikz}
    \usepackage{tikzscale}
    \usepackage{multirow}
    \usetikzlibrary{external}
    \usetikzlibrary{fixedpointarithmetic}
    \usetikzlibrary{spy}
    \usepackage{stfloats}
    \usepgfplotslibrary{groupplots}  
    \usepgfplotslibrary{colorbrewer}
    \pgfplotsset{filter discard warning=false}      
    \graphicspath{{../pdfs/}{../jpeg/}}
\else
\fi
\usepackage{amsmath}
\usepackage{amssymb}
\usepackage{mathtools}
\usepackage{siunitx}
\usepackage{bm}
\definecolor{purple}{rgb}{0,0,0}
\begin{document}
%
\title{Geometric Shaping of 2-Dimensional Constellations in the Presence of Laser Phase Noise}
%
%
%

\author{Hubert~Dzieciol,~\IEEEmembership{Student~Member,~IEEE},
        Gabriele~Liga,~\IEEEmembership{Member,~IEEE},
        Eric~Sillekens,~\IEEEmembership{Member,~IEEE},
        Polina~Bayvel,~\IEEEmembership{Fellow,~IEEE},
        and~Domani\c{c}~Lavery,~\IEEEmembership{Member,~IEEE}%

\thanks{Manuscript received May, 2020.
This work was funded by the United Kingdom (UK) Engineering and Physical Sciences Research Council (EPSRC) Programme Grant TRANSNET (Transforming networks - building an intelligent optical infrastructure, EP/R035342/1). H.~Dzieciol is in receipt of a PhD studentship from the EPSRC and Microsoft Research. The work of G. Liga is funded by the EUROTECH postdoc programme under the European Union’s Horizon 2020 research and innovation programme (Marie Skłodowska-Curie grant agreement No 754462). D.~Lavery is supported by the Royal Academy of Engineering under the Research Fellowships scheme.}
\thanks{H.~Dzieciol, E.~Sillekens, P.~Bayvel and D.~Lavery are with the Optical Networks Group, Department of Electronic and Electrical Engineering, University College London, London WC1E 7JE, U.K. (e-mail: \{hubert.dzieciol.18, e.sillekens, p.bayvel, d.lavery\}@ucl.ac.uk).}
\thanks{G.~Liga is with the Signal Processing Systems (SPS) Group, Department of Electrical Engineering, Eindhoven  University of Technology, The Netherlands. (e-mail: g.liga@tue.nl).}}

%
%

\markboth{PREPRINT}{}
%



\maketitle

\begin{abstract}
In this paper, we propose a geometric shaping (GS) strategy to design 8, 16, 32 and 64\textit{-ary} modulation formats for the optical fibre channel impaired by both additive white Gaussian (AWGN) and phase noise. The constellations were optimised to maximise generalised mutual information (GMI) using a mismatched channel model. The presented formats demonstrate an enhanced signal-to-noise ratio (SNR) tolerance in high phase noise regimes when compared with their quadrature amplitude modulation (QAM) or AWGN-optimised counterparts. By putting the optimisation results in the context of the 400ZR implementation agreement, we show that GS alone can either relax the laser linewidth (LW) or carrier phase estimation (CPE) requirements of 400~Gbit/s transmission links and beyond. Following the GMI validation, the performance of the presented formats was examined in terms of post forward error correction (FEC) bit-error-rate (BER) for a soft decision (SD) extended Hamming code (128, 120), implemented as per the 400ZR implementation agreement. We demonstrate gains of up to 1.2~dB when compared to the 64\textit{-ary} AWGN shaped formats. 
\end{abstract}

%
\IEEEpeerreviewmaketitle

\section{Introduction}
\label{sec:intro}
\IEEEPARstart{W}{hen} designing signals to communicate over optical fibres, one must consider all the sources of noise present in the communication channel. The mapping of digital information into a set of symbols is a key aspect of signal design that is sensitive to the statistics of the channel noise. The vast majority of research in this area has focused on the additive white Gaussian noise (AWGN) channel, however this is a sub-optimal approximation for all optical fibre communication systems, which include a range of other linear and nonlinear sources of noise. Specifically in this paper, we are concerned with the residual phase noise (RPN) present in a signal after coherent detection, which is caused by the finite laser linewidth (LW) of both the source and local oscillator (LO) lasers. 

Digital coherent communication systems use digital signal processing (DSP) to track the carrier phase. If the lasers used in such a system are sufficiently low LW then the residual phase noise after carrier phase estimation (CPE) can generally be ignored for the purposes of signal design. However, in low-complexity coherent communication systems, which use relatively high-LW lasers, significant signal-to-noise ratio (SNR) penalties can be incurred if the signal is not designed with respect to both AWGN and RPN. A current example of such a system is the 400ZR inter-data center standard which specifies laser LWs below 1~MHz with 60~GBd 16-\textit{ary} quadrature amplitude modulation (QAM) \cite{oif}.

The existing literature is abundant with works on geometrically designing modulation formats for enhanced sensitivity in a purely AWGN channel \cite{Foschini_AWGN_1974, Kernighan_Heur_1973, Forney_1984, codes_se}. A lower but still significant amount of work was also performed on channels with both phase noise and AWGN \cite{Hager_2013, Krishnan_2013, Moore_2009,Li_2008, Kayhan_2014, Sales_2019, Kwak_2008}. For instance, in the seminal work by Foschini in the 1970s \cite{Foschini_AWGNPN_1973}, an approximated model for the partially-coherent AWGN (PCAWGN) channel was introduced. This is a channel impaired by both AWGN and phase jitter. {\color{purple} Alternatively, the work in \cite{layton_2018} proposed a pilot-based demapper that, by incorporating the covariance information, captures the dependencies between the in-phase and quadrature components across the received constellation symbols in PCAWGN and nonlinear channel scenarios.}

In modern fibre optic communication systems, which pair high-order modulation with forward-error-correction (FEC) codes \cite{oif} (coded modulation, CM) the optimum choice of modulation format also depends on the SNR that can be tolerated at specific code rates. The most common implementation of CM, bit-interleaved CM (BICM)  \cite{szczecinski_alvarado_2015}, offers a straightforward method of bit-to-symbol mapping, effectively decoupling the coded bit sequence from the modulation format. In BICM systems, the so-called generalized mutual information (GMI) between input and output bits has been shown to be an accurate predictor of the post-FEC bit-error-rate (BER) performance \cite{Alvarado2016, Agrell_2018}. Moreover, the GMI was shown to be an achievable information rate (AIR) for BICM systems. Consequently, recent work on geometric shaping has used GMI as a metric to optimise modulation formats in the presence of a range of channel impairments, including AWGN, nonlinear interference, and peak-to-average power ratio transmitter constraints \cite{Eric_2018, Ionescu_2019, Chen_2018_GS64, Gerard_2019, Zhang_2017, Millar_2018} .

Even though geometric shaping (GS) for the AWGN channel has been shown to increase the AIRs in optical fibre systems \cite{Eric_2018, Ionescu_2019, Chen_2018_GS64, Gerard_2019}, few have addressed the problem of RPN. The work presented in \cite{Pfau_2011} focused on 16-point constellations for hard-decision, symbol-wise decoders. The work in \cite{Hager_2013} addressed the nonlinear impairments in long-haul transmission using symbol error probability as the performance indicator, whereas in \cite{Krishnan_2013} the modulation formats were optimised to maximise the mutual information (MI). There are several examples of modified gradient descent \cite{Gerard_2019} and pairwise optimisation techniques \cite{Pfau_2011, Chen_2018_GS64, Moore_2009} that have shown increased transmission rates compared to the conventional QAM formats in the AWGN channel. Moreover, various investigations from other areas of communications have shown considerable gains from jointly optimising constellations for both AWGN and phase noise 
\cite{Li_2008}. For example, \cite{Kayhan_2014} optimised the constellations for GMI. 
\begin{figure*}[t]
\centering
\includegraphics[width=0.9\linewidth]{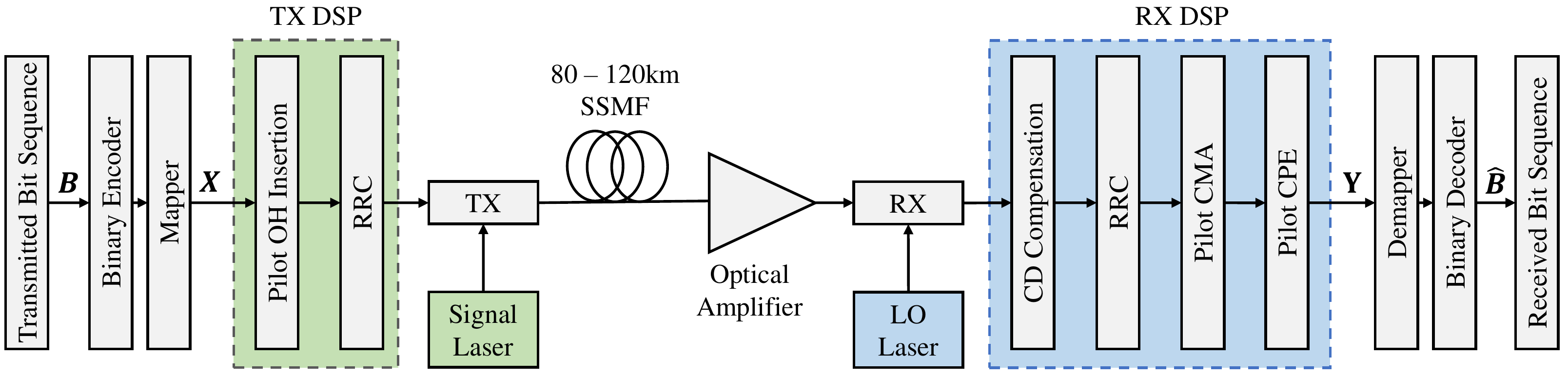}
\caption{A typical short-range optical fibre communication link employing BICM with soft decision FEC (SD-FEC).}
\label{fig:system_model}
\vspace{-12pt}
\end{figure*}
Another important parameter in signal design is the channel model. A common detection strategy, applied throughout many areas of communication, is based on mismatched reception \cite{Merhav_1994}. Under the assumption of a circular Gaussian noise distribution, it uses the minimum Euclidean distance between the expected and received data to make symbol-or bit-wise decisions. However, in the presence of both AWGN and phase noise components, the Gaussian noise assumption is no longer accurate. More sophisticated channel models have been proposed \cite{Foschini_AWGNPN_1973, Sales_2019} and, in some cases\cite{Kayhan_2014}, used for GS. However, when compared with the mismatched reception, they often require much higher implementation complexity \cite[Table 1]{Sales_2019}, which has a direct impact on the complexity of any GS optimisation algorithm based on the improved channel model. 
%

This paper provides a fresh perspective on the design of GS modulation formats for optical fibre communication systems to reduce the implementation complexity. The indicative system parameters are chosen with a reference to the 400ZR implementation agreement \cite{oif}, which specifies 16-\textit{ary} modulation, although we extend the scope of this work by additional investigation of 8, 32 and 64-\textit{ary} formats and a range of laser LWs to go beyond these industrial benchmark specifications. 

The remainder of this document is organised as follows: the proposed system model is described in the Section \ref{sec:sysmodel}. Subsequently, Section \ref{sec:cmodulation} introduces the auxiliary channel model employed to optimise the constellations presented in this paper. Additionally, it provides an overview of the information theoretic metrics that are commonly applied to approximate the performance of BICM driven schemes. Section \ref{sec:methodology} contains a detailed description of the shaping methodology and the verification method used to validate the results presented in Sections \ref{sec:results} and \ref{sec:fec_results}. The former section presents the GMI performance of the constellations optimised for both AWGN and phase noise, whereas the latter examines the post-FEC BER of each format. Ultimately, Section \ref{sec:conclusion} summarises the findings; drawing a general conclusion and outlines the potential directions for future research.
\section{System Model}
\label{sec:sysmodel}
\begin{figure}[t]
    \centering
    \begin{minipage}{\columnwidth}
        \centering
        \scalebox{1.0}{
        \includegraphics[width=0.9\linewidth]{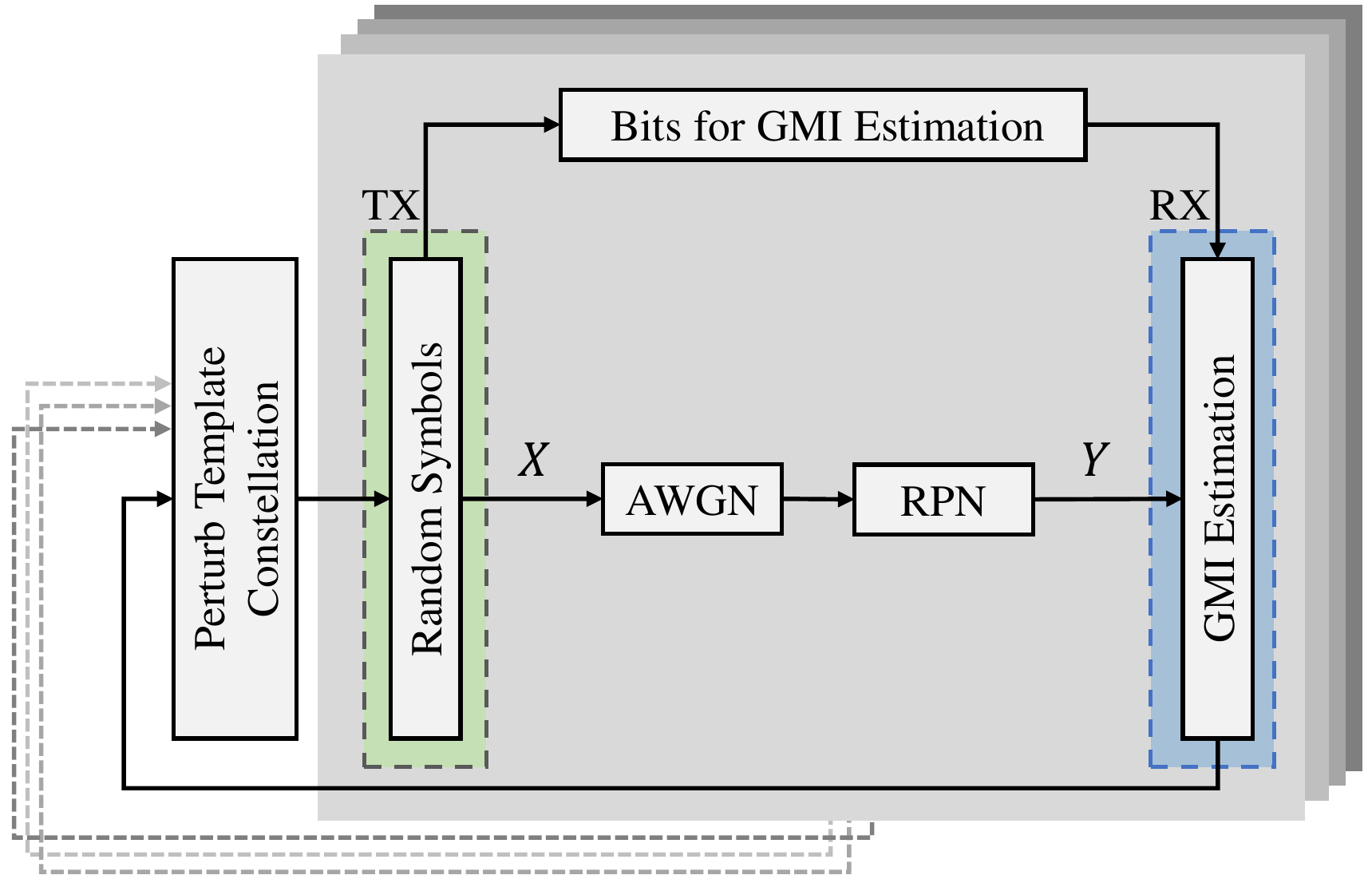}}\hfill
    \end{minipage}%

\caption{A visual representation of the shaping algorithm, where $X$ and $Y$ stand for the originally transmitted and the received symbols, respectively. The gray layers indicate a multi-dimensional characteristic of the optimiser, i.e. the perturb template constellation is updated individually for every symbol coordinate change. The most optimal arrangement out of all $2 \times M$-symbol combinations is selected at each iteration until the algorithm converges to a local optimum.}
\label{fig:shaping_algo_model}
\vspace{-12pt}
\end{figure}
\begin{figure}[t]
    \centering
    \begin{minipage}{\columnwidth}
        \centering
        \scalebox{1.0}{
        \includegraphics[width=0.9\linewidth]{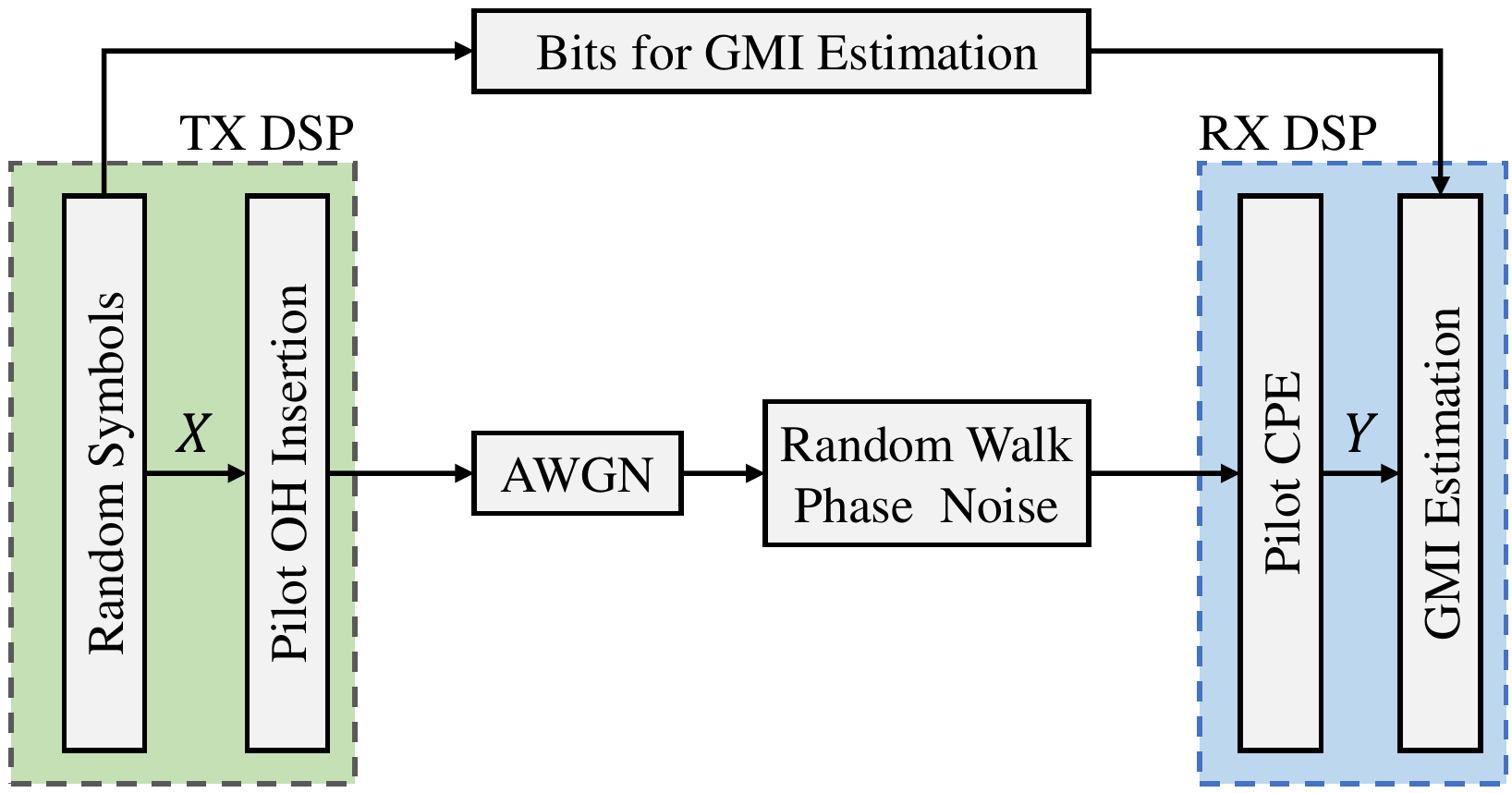}}\hfill
    \end{minipage}%

\caption{Verification model of the presented constellations, where $X$ and $Y$ represent the originally transmitted and the received symbols, respectively. The performance is estimated via GMI calculation with respect to the originally transmitted random bit sequence.}
\label{fig:verification_model}
\vspace{-12pt}
\end{figure}
\raggedbottom

\begin{figure}[!ht]
    \centering
	\includegraphics[height=0.29\textwidth]{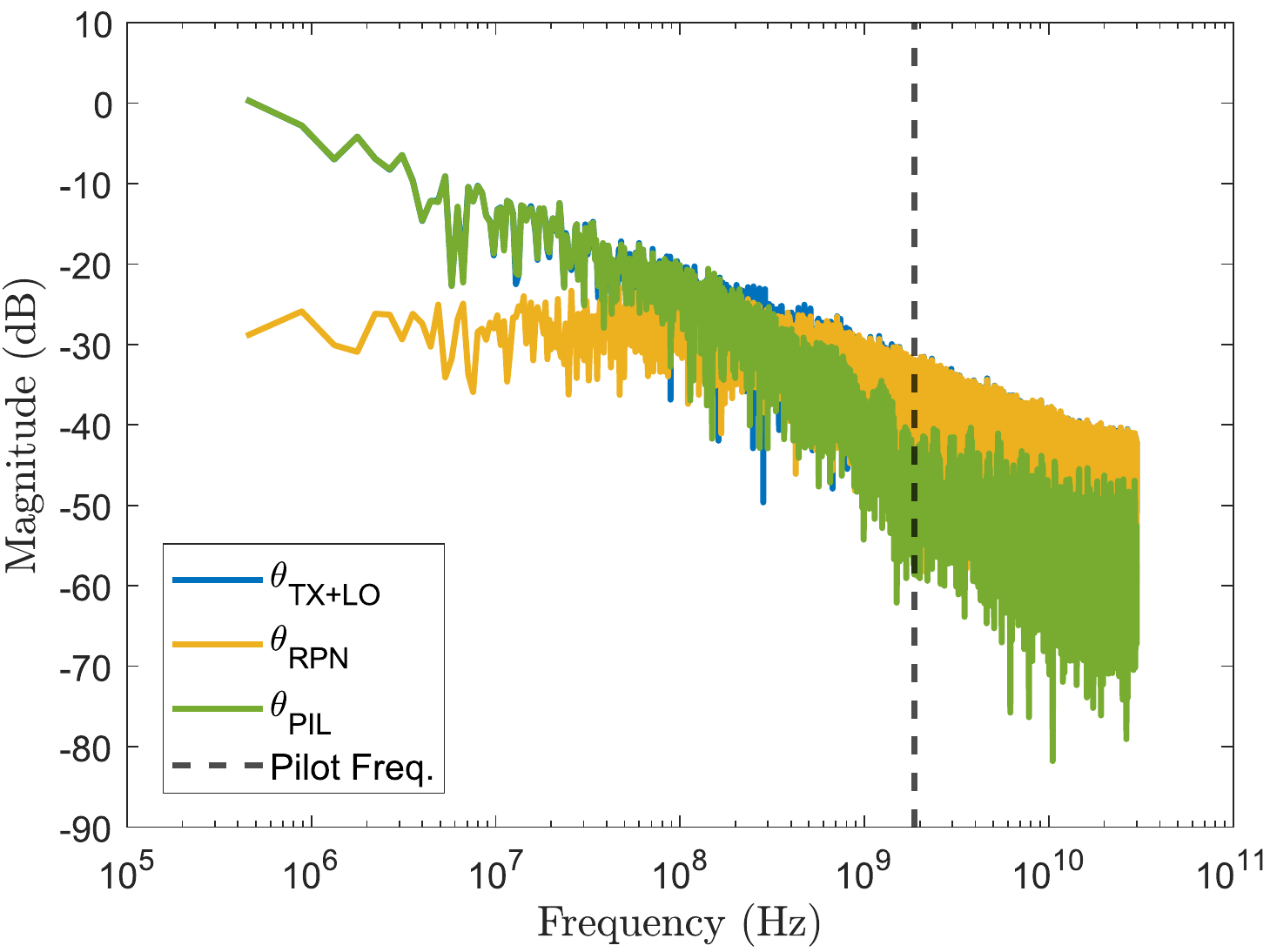}
    \caption{Typical frequency spectra of the phase noise contributions in the presented system.}
    \label{fig:phase_fs}
    \vspace{-12pt}
\end{figure}
Fig. \ref{fig:system_model} depicts a typical implementation of a short-range optical fibre link. Such a model addresses the requirements of 400ZR \cite{oif} and serves as a reference system for the results presented in this paper. {\color{purple}In the following configuration, a binary FEC encoder adds redundancy to a randomly generated uncoded bit sequence $\boldsymbol{B}$ and maps it into a sequence of $N$ symbols $\boldsymbol{X}=[X_1,X_2,...X_N]$ each drawn from a complex constellation $\mathcal{C} =\{C_{1}, C_{2}, \ldots, C_{M}\}$ of size $M$.} Subsequently, QPSK symbols are generated as a pilot overhead (OH), as per the 400ZR agreement, which are embedded into the data frame. A pilot is inserted after every block of 32 data symbols. Then, the signal is pulse-shaped using a root-raised cosine (RRC) filter.

In the next stage, the shaped pulses are ideally converted into the optical domain by the transmitter (TX) and, as per 400ZR specification, carried over $80$--$120$~km of a standard single-mode fibre (SSMF). The incoming single-channel signal is then optically amplified and ideally converted into the digital domain with an optical pass-band filter and a coherent receiver (RX). To recover the carrier phase information, the signal is mixed with a local reference laser (LO). Then, converted to a digital domain to remove the channel impairments with the sufficient DSP techniques.\begin{figure*}[!b]
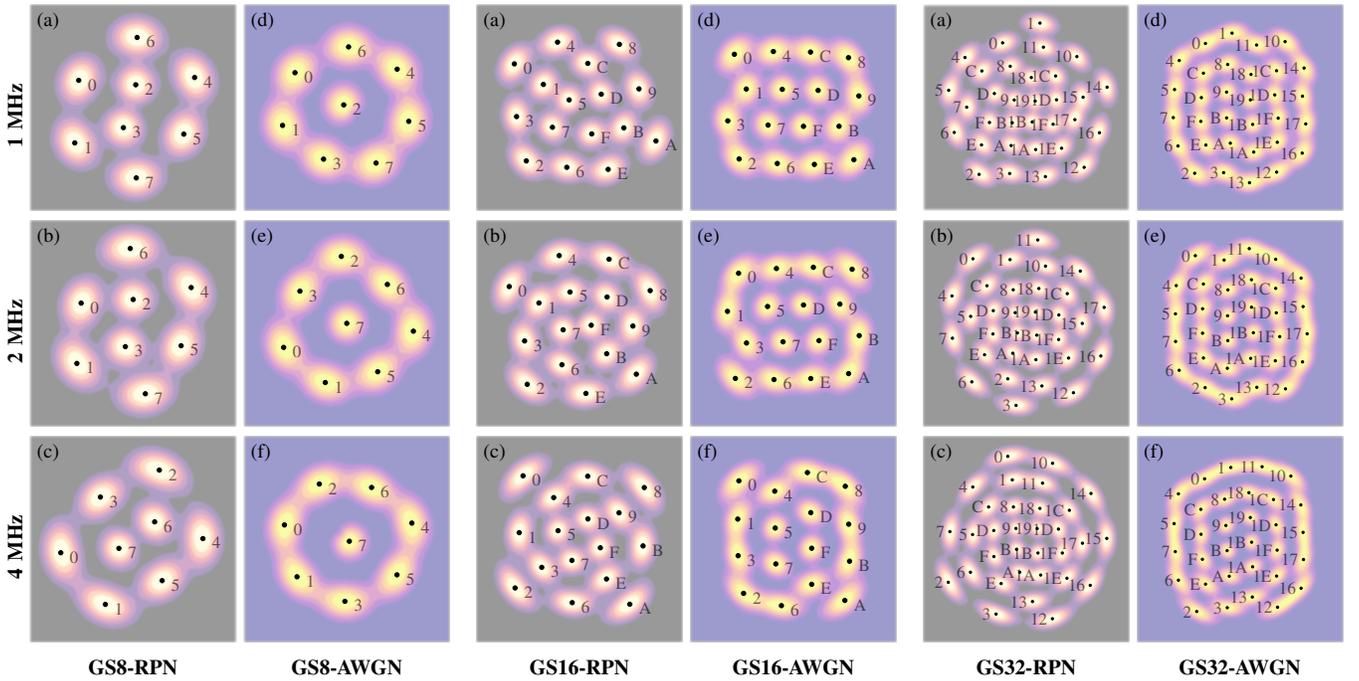

\centering

\begin{tabular}{c@{\hspace{.1cm}}c@{\hspace{.1cm}}c @{\hspace{.35cm}}c@{\hspace{.1cm}}c@{\hspace{.35cm}}c@{\hspace{.1cm}}c}
\rotatebox{90}{\makebox[1in][c]{\footnotesize{\textbf{1~MHz}}}} &
\includegraphics[width=1.08in]{8QAM/Mis/LW=1.0__SNR=11.5dB_RPN_MIS.tikz} &
\includegraphics[width=1.08in]{8QAM/Mat/LW=1.0__SNR=11.5dB_AWGN_MAT.tikz} &
\includegraphics[width=1.08in]{16QAM/Mis/LW=1.0__SNR=14.5dB_RPN_MIS.tikz} &
\includegraphics[width=1.08in]{16QAM/Mat/LW=1.0__SNR=14.5dB_AWGN_MAT.tikz} &
\includegraphics[width=1.08in]{32QAM/Mis/LW=1.0__SNR=17.5dB_RPN_MIS.tikz} &
\includegraphics[width=1.08in]{32QAM/Mat/LW=1.0__SNR=17.5dB_AWGN_MAT.tikz} \\
\rotatebox{90}{\makebox[1in][c]{\footnotesize{\textbf{2~MHz}}}} &
\includegraphics[width=1.08in]{8QAM/Mis/LW=2.0__SNR=11.5dB_RPN_MIS.tikz} &
\includegraphics[width=1.08in]{8QAM/Mat/LW=2.0__SNR=12.0dB_AWGN_MAT.tikz} &
\includegraphics[width=1.08in]{16QAM/Mis/LW=2.0__SNR=14.5dB_RPN_MIS.tikz} &
\includegraphics[width=1.08in]{16QAM/Mat/LW=2.0__SNR=15.0dB_AWGN_MAT.tikz} &
\includegraphics[width=1.08in]{32QAM/Mis/LW=2.0__SNR=18.0dB_RPN_MIS.tikz} &
\includegraphics[width=1.08in]{32QAM/Mat/LW=2.0__SNR=18.0dB_AWGN_MAT.tikz} \\
\rotatebox{90}{\makebox[1in][c]{\footnotesize{\textbf{4~MHz}}}} &
\includegraphics[width=1.08in]{8QAM/Mis/LW=4.0__SNR=12.0dB_RPN_MIS.tikz} &
\includegraphics[width=1.08in]{8QAM/Mat/LW=4.0__SNR=12.5dB_AWGN_MAT.tikz} &
\includegraphics[width=1.08in]{16QAM/Mis/LW=4.0__SNR=15.0dB_RPN_MIS.tikz} &
\includegraphics[width=1.08in]{16QAM/Mat/LW=4.0__SNR=15.5dB_AWGN_MAT.tikz} &
\includegraphics[width=1.08in]{32QAM/Mis/LW=4.0__SNR=18.5dB_RPN_MIS.tikz} &
\includegraphics[width=1.08in]{32QAM/Mat/LW=4.0__SNR=19.0dB_AWGN_MAT.tikz} \\
&
\footnotesize{\textbf{GS8-RPN}} &
\footnotesize{\textbf{GS8-AWGN}} &
\footnotesize{\textbf{GS16-RPN}} &
\footnotesize{\textbf{GS16-AWGN}} &
\footnotesize{\textbf{GS32-RPN}} &
\footnotesize{\textbf{GS32-AWGN}} \\
\end{tabular}

\caption{The RPN shaped (GS$M$-RPN) 8-\textit{ary}, 16-\textit{ary} and 32-\textit{ary} constellations for: (a) $1$~MHz, (b) $2$~MHz and (c) $4$~MHz laser linewidths. Constellations (d), (e) and (f) are the GS$M$-AWGN counterparts for $1$~MHz, $2$~MHz and $4$~MHz, respectively. All the formats depict optimal symbol arrangement at the indicative GMI threshold of 0.96$\times{}$m~bit/symbol at $60$~Gbaud, bit-to-symbol mapping is given as hexadecimal numbers. Note that the 0.96$\times{}$m GMI threshold is used throughout this paper, and is discussed in section~\ref{sec:fec_results}}
\label{fig:jointccs}
\end{figure*}
Following the chromatic dispersion (CD) compensation and matched filtering, the received data is equalized with a pilot-based algorithm \cite{Mazur_19, Yuta_2019}. 
The resulting sequence of received symbols $\boldsymbol{Y}=[Y_1,Y_2,...Y_N]$ is then demapped and subsequently decoded to obtain a sequence of estimated symbols $\boldsymbol{\hat{B}}$.

{\color{purple}
In a coherent detection scheme, the phase information of a received optical signal $\theta _{\text{TX}}$ is mixed with the phase of a LO laser $\theta_{\text{LO}}$ which, in principle, is tuned closely to the transmitter's optical  wavelength. Fig.~\ref{fig:phase_fs} depicts typical frequency spectra of the phase noise contributions in the presented system. At this stage, the phase of the received signal can be modelled as a ``random-walk" process $\theta_{\text{TX+LO}}$ (blue line). Subsequently, a portion of the carrier phase information $\theta_{\text{PIL}}$ is estimated by the CPE. In the presented system, the carrier phase is estimated by applying a Wiener filter to the QPSK pilots that are embedded in-between the data symbols (further explained in Sec.~\ref{sec:methodology}-B). The resulting phase estimates are denoted as  $\theta _{\text{PIL}}$ (green line). Consequently, by subtracting $\theta_{\text{PIL}}$ from $\theta_{\text{TX+LO}}$, we end up with $\theta_{\text{RPN}}$ (orange line), which is the remaining portion of the phase noise that is left in the signal after the CPE. In the remainder of this paper, we refer to it as the RPN. Hence, at symbol period $k$, a discrete filter output of the carrier phase recovery operating at the symbol rate, can be described as
\begin{equation}\label{eq:RPN}
    \theta _{\text{RPN}}(k) = \theta _{\text{TX+LO}}(k) - \theta _{\text{PIL}}(k).
\end{equation}
%

To optimise the constellations more efficiently (described in Sec.~\ref{sec:methodology}-A.), we approximated the RPN contribution (in the optimiser) as a random, zero-mean Gaussian distributed variable with a variance
\begin{equation}\label{eq:rpn_lw}
\sigma^2_{\text{RPN}}=2\pi\Delta\nu T_{s},
\end{equation}
where $T_{s}$ denotes the symbol period and $\Delta\nu$ is a combined linewidth of the LO and the TX laser\cite{silfvast_2009}.}
It should be highlighted that the constellations presented in this paper are optimised for the residual phase noise (RPN) after the CPE. Therefore, a ``random-walk" component is assumed to be compensated for at this stage. Hence, the resulting discrete-time channel relationship can be expressed as
\begin{equation}
\label{eq:chm}
Y_k = X_k \cdot e^{j\theta _{\text{RPN}}(k)} + \mathcal{N}_k, \;\; \text{for} \;k=1,2,...,N, 
\end{equation}
where $\mathcal{N}_k$ represents a complex zero-mean Gaussian random variable. Following the compensation of the ``random-walk" component of the phase noise, a BICM demapper passes the soft information about the recovered symbols $\boldsymbol{Y}$ to a binary decoder to recover the data bits $\widehat{\bm{B}}$.
\section{Coded Modulation}
\label{sec:cmodulation}
\begin{figure*}[!b]
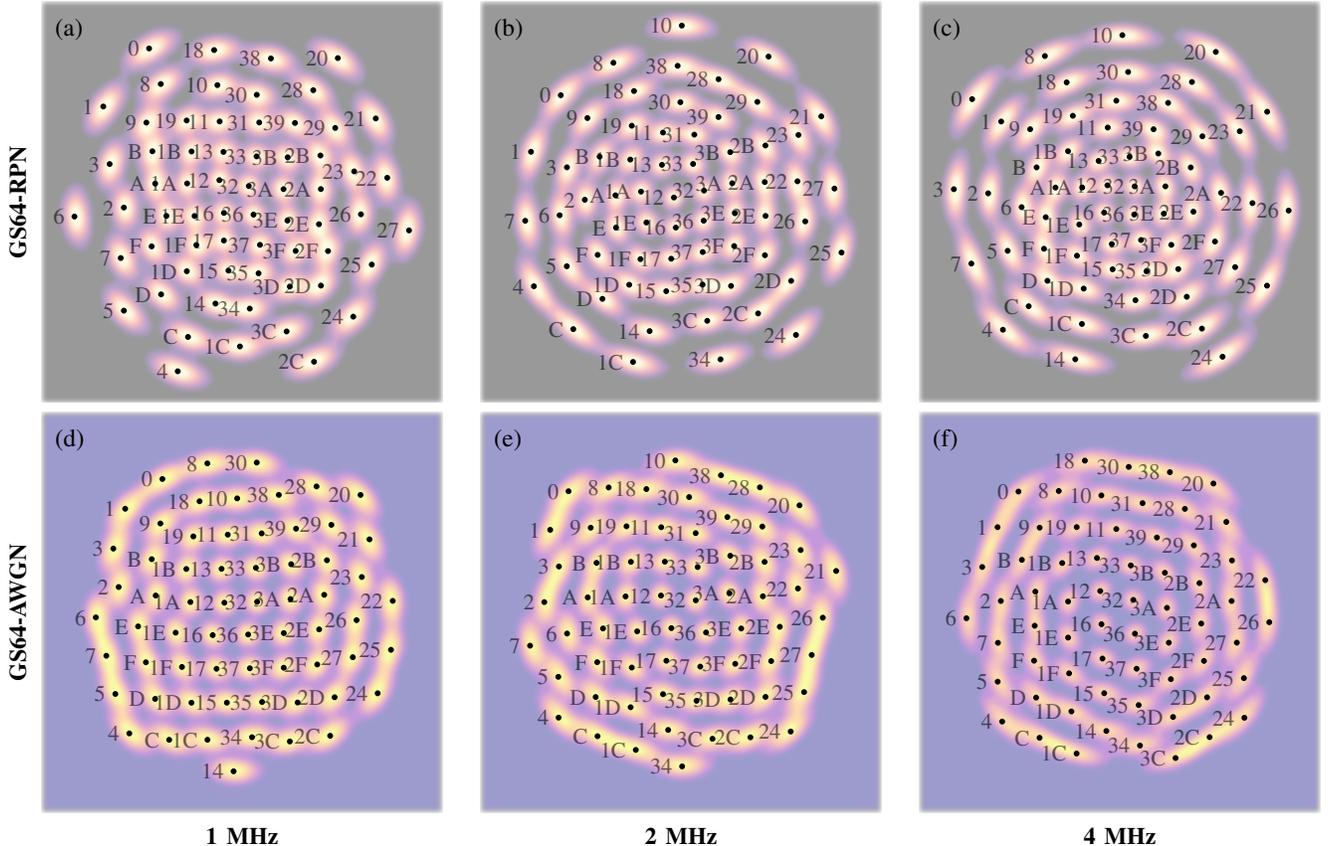

\centering
\begin{tabular}{c@{\hspace{.2cm}}c@{\hspace{.5cm}}c@{\hspace{.5cm}}c}
\rotatebox{90}{\makebox[2.1in][c]{\small{\textbf{GS64-RPN}}}} &
\includegraphics[width=2.1in]{64QAM/Mis/LW=1.0__SNR=20.5dB_RPN_MIS.tikz} &
\includegraphics[width=2.1in]{64QAM/Mis/LW=2.0__SNR=21.5dB_RPN_MIS.tikz} &
\includegraphics[width=2.1in]{64QAM/Mis/LW=4.0__SNR=22.0dB_RPN_MIS.tikz} \\
\rotatebox{90}{\makebox[2.1in][c]{\small{\textbf{GS64-AWGN}}}} &
\includegraphics[width=2.1in]{64QAM/Mat/LW=1.0__SNR=21.0dB_AWGN_MAT.tikz} &
\includegraphics[width=2.1in]{64QAM/Mat/LW=2.0__SNR=21.5dB_AWGN_MAT.tikz} &
\includegraphics[width=2.1in]{64QAM/Mat/LW=4.0__SNR=22.5dB_AWGN_MAT.tikz} \\
&
\small{\textbf{1~MHz}} &
\small{\textbf{2~MHz}} &
\small{\textbf{4~MHz}} \\
\end{tabular}
\caption{Phase noise shaped 64-\textit{ary} constellations for: (a) $1$~MHz, (b) $2$~MHz and (c) $4$~MHz. Constellations (d), (e) and (f) are the GS64-AWGN counterparts for $1$~MHz, $2$~MHz and $4$~MHz, respectively. All the formats depict optimal symbol arrangement at the indicative GMI threshold of 0.96$\times{}$m~bit/symbol at $60$~Gbaud, bit-to-symbol mapping is given as hexadecimal numbers.}
\label{fig:64qam_cc}
\vspace{-12pt}
\end{figure*}
\subsection{AWGN auxiliary channel model}
The demapper computes soft information about the transmitted bits assuming a channel law which is expected to closely approximate the actual channel. Nevertheless, computing such quantities can be computationally prohibitive \cite[Table 1]{Sales_2019}. Therefore, a more common approach in fibre optics is to use a mismatched Gaussian receiver \cite{Merhav_1994, Bocherer_2019}, using the channel law
\begin{equation}
f_{Y | X}(y | x)=\frac{1}{\left(\pi N_{0}\right)^{2}} \exp \left(-\frac{\|y-x\|^{2}}{N_{0}}\right),
\end{equation}
where the temporal index $n$ is here dropped in the generic symbols $X_k$ and $Y_k$ due to the underlying stationarity assumption, and $N_{0}$ is a 2D variance of an independent, zero-mean, random Gaussian variable vector and the $||y  - x||^2$ denotes a squared Euclidean distance between the symbols.

\subsection{Performance metrics} Although it is in principle possible to optimize the geometry of a modulation format in terms of post-FEC BER for each specific FEC scheme, this would be highly impractical. Conveniently, information-theoretic metrics allow an accurate prediction of the system performance for a large class of good FEC codes. Considering a non-binary SD-FEC encoder-decoder pair, MI is the most accurate performance predictor \cite{Schmalen2017}. For any memoryless optical fibre channel, it can be formulated as
\begin{equation}
\mathrm{MI}=I(X ; Y) \triangleq \mathbb{E}\left[\log _{2} \frac{f_{Y | X}(Y | X)}{f_{Y}(Y)}\right],
\end{equation}
where $X$ stands for the input, $Y$ represents the output symbols, $\mathbb{E}$ is the expectation and $f_{Y | X}$ denotes the channel law. In case of binary SD-FEC codes, the performance can be predicted via the GMI \cite{Kaplan_1993, Alvarado2016} which is defined as{\color{purple} 
%
\begin{equation}
\mathrm{GMI} \triangleq I\left(B_{i} ; Y\right)=  \max_{s \geq 0}  \sum_{i=1}^{m} \mathbb{E}\left[\log _{2} \frac{f_{Y | B_{i}}\left(Y | B_{i}\right)^{s}}{f_{Y}(Y)^{s}}\right],
\end{equation}
}
where $M$ is the constellation order, $m = \log_2M$ is the number of bits mapped into each trasmitted symbol $X$. In this work, we will focus on the GMI as a metric to maximise the performance of modulation formats in the RPN channel when soft-information is passed to the FEC decoder (see Sec.~\ref{sec:fec_results}). 

{\color{purple} In the remainder of this manuscript, we set the term $s$ to $s=1$ as we didn't observe any significant increase of the achieved information rates by adding this additional degree of freedom to the shaping algorithm. The evidence suggests that scaling the log-likelihood ratios has a negligible benefit in the context of the presented system specifications.} 

\section{Methodology}
\label{sec:methodology}
\begin{figure*}[!t]
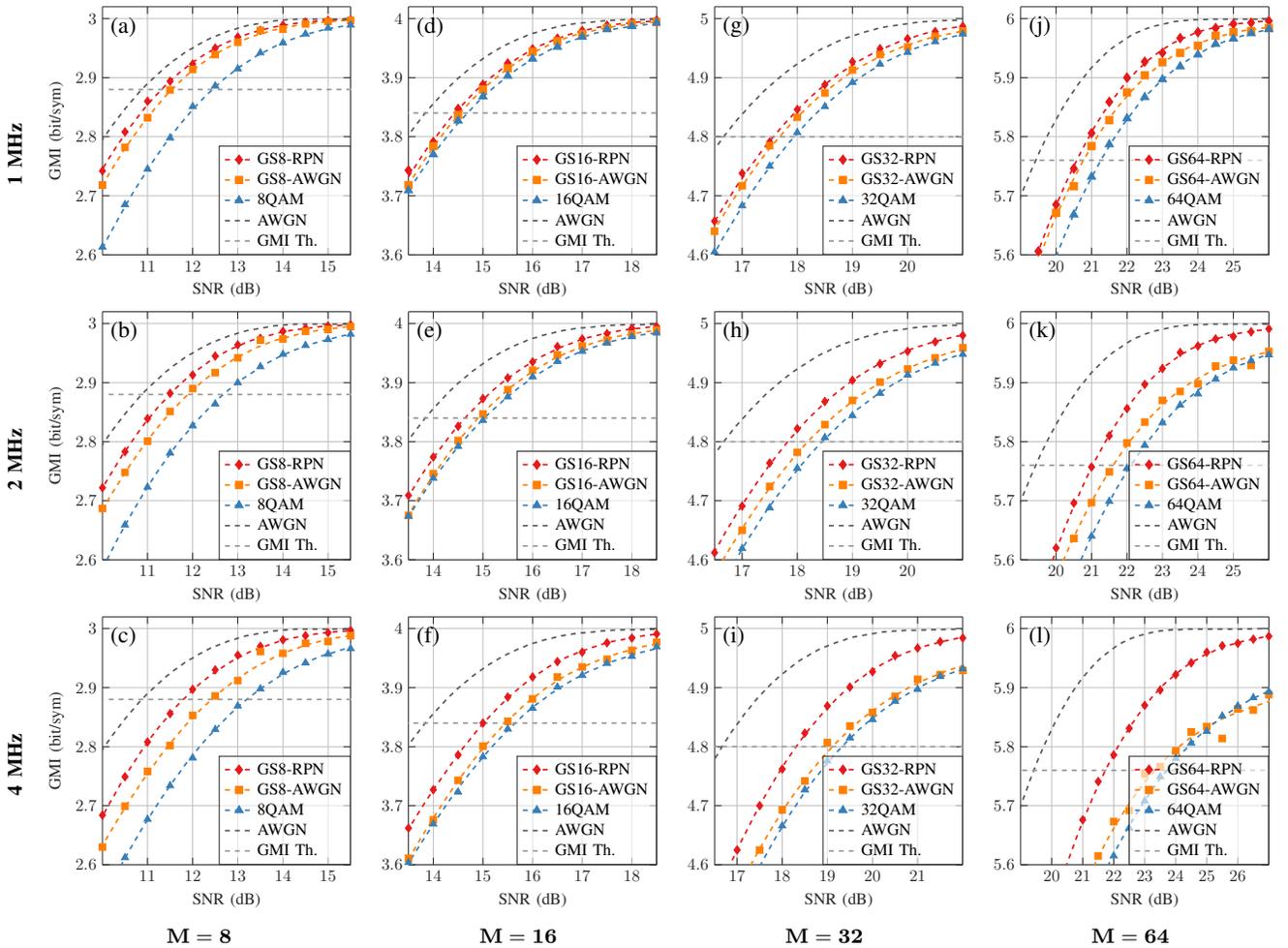

\centering

\begin{tabular}{c@{\hspace{.3cm}}c@{\hspace{.3cm}}c @{\hspace{.3cm}}c@{\hspace{.3cm}}c}
\rotatebox{90}{\makebox[1.6in][c]{\footnotesize{\textbf{1~MHz}}}} &
\includegraphics[height=1.6in]{8QAM/Curves/GMI_1MHz.tikz} &
\includegraphics[height=1.6in]{16QAM/Curves/GMI_1MHz.tikz} &
\includegraphics[height=1.6in]{32QAM/Curves/GMI_1MHz.tikz} &
\includegraphics[height=1.6in]{64QAM/Curves/GMI_1MHz.tikz} \\
\rotatebox{90}{\makebox[1.6in][c]{\footnotesize{\textbf{2~MHz}}}} &
\includegraphics[height=1.6in]{8QAM/Curves/GMI_2MHz.tikz} &
\includegraphics[height=1.6in]{16QAM/Curves/GMI_2MHz.tikz} &
\includegraphics[height=1.6in]{32QAM/Curves/GMI_2MHz.tikz} &
\includegraphics[height=1.6in]{64QAM/Curves/GMI_2MHz.tikz} \\
\rotatebox{90}{\makebox[1.6in][c]{\footnotesize{\textbf{4~MHz}}}} &
\includegraphics[height=1.6in]{8QAM/Curves/GMI_4MHz.tikz} &
\includegraphics[height=1.6in]{16QAM/Curves/GMI_4MHz.tikz} &
\includegraphics[height=1.6in]{32QAM/Curves/GMI_4MHz.tikz} &
\includegraphics[height=1.6in]{64QAM/Curves/GMI_4MHz.tikz} \\
&
\footnotesize{$\mathbf{M=8}$} &
\footnotesize{$\mathbf{M=16}$} &
\footnotesize{$\mathbf{M=32}$} &
\footnotesize{$\mathbf{M=64}$} \\
\end{tabular}

\caption{GMI curves for: 8-\textit{ary} (a-c), 16-\textit{ary} (d-f), 32-\textit{ary} (g-i) and 64-\textit{ary} (j-l) constellations at 60~Gbaud, single-channel transmission. Modulation orders ($M$) are grouped in columns, whereas each row represents the combined laser linewidths of 1~MHz, 2~MHz and 4~MHz, respectively. A GMI threshold of 0.96$\times{}$m bit/symbol used for the results in Figs.~\ref{fig:shaping_gain}, \ref{fig:curves_awgn} and \ref{fig:coding_gain}, is also shown.}
\label{fig:gmi_grid}
\vspace{-12pt}
\end{figure*}
\subsection{Constellation optimisation}\label{subsec:constopt}
The shaping technique adopted in this work, is schematically illustrated in Fig.~\ref{fig:shaping_algo_model}. It combines the first stage of the gradient descent algorithm from \cite{Schrekenbach_2013} with a binary label switching method from \cite{Chen_2018_GS64} to optimise bit-to-symbol mapping. {\color{purple} Consequently, it aims to maximise the GMI performance of the modulation format.} It should be noted that this approach is not guaranteed to find the global optimum. However, the resulting constellations outperformed their $M$QAM counterparts in all cases.

The constellations which were shaped assuming an AWGN channel, herein referred to as GS$M$-AWGN, were optimised using Gauss-Hermite numerical integration formula \cite{salzer_1952} to compute the GMI. This case corresponds to the channel model in (\ref{eq:chm}), where the phase noise $\theta_{\text{RPN}}(k)$ is identically zero. Subsequently, constellations were optimised for a channel consisting of both AWGN and RPN components. For this, the GS$M$-AWGN formats were used to initialise the Monte-Carlo optimiser with $2^{19}$ randomly interleaved symbols per iteration. The resulting constellations are referred to as GS$M$-RPN.

All presented formats were individually optimised over a range of SNR and RPN combinations, then evaluated in the context of the 400ZR specifications. Hence, the investigated RPN variances correspond to the $1$~MHz and $2$~MHz combined laser LWs at the transmission rate of $60$~Gbaud. Additionally, we extended the scope of the investigation with a $4$~MHz configuration to investigate the potential benefits of GS$M$-RPN shaping in higher phase noise regimes.

\subsection{Validation of the optimized constellations}\label{subsec:constvalid}
In the previous section, the GMI optimization was performed using a phase-noise channel where the LW was set directly to the targeted value of RPN. However, as described in \eqref{eq:RPN}, the RPN is the result of an imperfect CPE when using receiver DSP. Thus, in order to validate the shaped constellations obtained under the procedure in  Sec.~\ref{subsec:constopt} in such a scenario, we used the approach depicted in Fig.~\ref{fig:verification_model}.
The schematic diagram in Fig.~\ref{fig:verification_model} shows the Monte-Carlo GMI estimation in an RPN channel consisting of AWGN and a random walk phase-noise process followed by a pilot-aided CPE. {\color{purple} For each set of channel parameters (SNR and RPN LW), a sequence of $2^{20}$ random symbols was used simulated.} The CPE was undertaken with a format-agnostic recovery scheme \cite{Mazur_19, Yuta_2019}, where the QPSK pilots were inserted after each block of 32 data symbols. Then, the estimated phase information was interpolated over the entire data frame using Wiener filter coefficients calculated from \cite{Ip_2007}. Subsequently, the phase information was estimated across at least two consecutive pilots to ensure correct interpolation. Hence, a minimum CPE filter length of 65 taps was used. 

Although using blind phase tracking between the pilot symbols could potentially enhance the CPE, the results presented herein can be viewed as a conservative estimate of the overall performance. (We note that a blind CPE is not specified in 400ZR.) Moreover, the phase noise component was modelled to produce approximately the same RPN variance as used for the corresponding GS-RPN optimisation.

{\color{purple} For all GMI calculations presented herein, note that the pilot overhead has not been included in the calculation. Explicitly this mean that the maximum GMI achievable for a constellation of order $M$ is $\log_2{(M)}$. It is trivial to calculate the GMI net of the pilot rate by multiplying the GMI by $32/33$.}
{\color{purple} Moreover, all the reported results assume that the pilot power is set to be equal to  the average signal power. By adopting the following approach, the CPE overhead has no impact on the SNR.}

\section{GS Results}
\label{sec:results}
\raggedbottom
In this Section, we present the GMI results for the GS$M$-RPN and GS$M$-AWGN
that were computed with the approach described in Sec.~\ref{subsec:constvalid}. Moreover, to provide a fairer comparison between the GS$M$-AWGN and GS$M$-RPN formats, we also compared their performance to the conventional $M$QAM constellations with Gray bit-to-symbol mapping.
\begin{figure*}[!t]
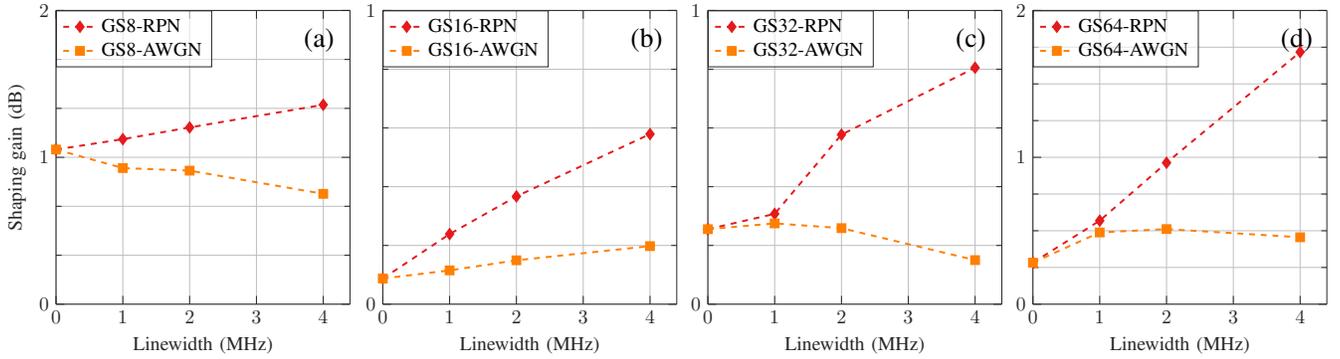

\centering
\begin{tabular}{c@{\hspace{.1cm}}c@{\hspace{.1cm}}c@{\hspace{.1cm}}c}
\includegraphics[height=1.9in]{8QAM/Curves/sh_gain.tikz}
\includegraphics[height=1.9in]{16QAM/Curves/sh_gain.tikz} &
\includegraphics[height=1.9in]{32QAM/Curves/sh_gain.tikz} &
\includegraphics[height=1.9in]{64QAM/Curves/sh_gain.tikz} \hfill\\
\end{tabular}
\caption{Shaping gains with respect to $M$QAM at the GMI=0.96$\times{}m$~bit/symbol at $60$~Gbaud, for: 8-\textit{ary} (a), 16-\textit{ary} (b), 32-\textit{ary} (c) and 64-\textit{ary} (d) constellations, respectively.}
\label{fig:shaping_gain}
\vspace{-12pt}
\end{figure*}

%

Figs.~\ref{fig:jointccs} and \ref{fig:64qam_cc} depict the 8, 16, 32 and 64-\textit{ary} formats optimized at a GMI of 0.96$\times{}m$~bit/symbol. The multi-layered background of each constellation provides a visual representation of the PCAWGN channel. It was generated with the approximation from \cite{Sales_2019}, whereas the symbol probability is proportional to the colour brightness. The orange/violet-coloured heatmap is used for GS$M$-AWGN, whereas the red/purple represents the GS$M$-RPN formats. In Fig.~\ref{fig:jointccs}, each row corresponds to a separate laser LW configuration, being 1~MHz, 2~MHz and 4~MHz, respectively, whereas each column stands for the corresponding modulation format. Additionally, the formats of the same cardinality are visually closer. Thus, forming the groups of six constellations each. To improve the readability of the particular bit mappings, Fig. \ref{fig:64qam_cc} depicts 64-\textit{ary} symbol arrangements, as being the most dense, in an inverted fashion. Consequently, the GS64-AWGN and GS64-RPN constellations are presented in two separate rows, whereas the corresponding laser LWs are denoted by columns. Compared to GS$M$-AWGN, it can be noticed that the GS$M$-RPN constellations tend to have more {\color{purple}circularly symmetric} shapes. Moreover, the symbols are grouped in straight lines originating from the center. Interestingly, the following behaviour can especially be observed in larger phase noise regimes of the higher cardinality configurations.

The curves illustrated in Fig.~\ref{fig:gmi_grid} show the GMI vs. SNR performance of GS$M$-RPN, GS$M$-AWGN, and $M$-QAM formats, for $M=8,16,32,64$, and LW=1 MHz, 2 MHz, and 4 MHz. The GS constellations are individually optimised for each SNR-RPN pair and the performance of GS$M$-AWGN in the AWGN channel (dashed black line) is also shown as a reference.

{\color{purple}
\begin{table}[!h]
\renewcommand\arraystretch{1.5}%
\centering
\caption{8-point format optimisation\tablefootnote{Shown as the shaping gains of GS8-RPN constellations with respect to the corresponding GS8-AWGN and 8QAM formats.}}
\begin{tabular}{cccc} 
\toprule
\begin{tabular}[c]{@{}c@{}}\textbf{LW (MHz)}\end{tabular} & \begin{tabular}[c]{@{}c@{}}\textbf{GS8-AWGN (dB)}\end{tabular} & \begin{tabular}[c]{@{}c@{}}\textbf{8QAM (dB)}\end{tabular} & \textbf{Fig.}  \\ 
\midrule
1 & 0.2 & 1.1 & \ref{fig:gmi_grid}(a) \\ 
2 & 0.3 & 1.2 & \ref{fig:gmi_grid}(b) \\ 
3 & 0.7 & 1.5 & \ref{fig:gmi_grid}(c) \\
\bottomrule
\end{tabular}
\label{tab:sg8}
\end{table}

\begin{table}[!h]
\renewcommand\arraystretch{1.5}%
\centering
\caption{16-point format optimisation\tablefootnote{Shown as the approximated shaping gains of GS16-RPN constellations with respect to the corresponding GS16-AWGN and 16QAM formats.}}
\begin{tabular}{cccc} 
\toprule
\begin{tabular}[c]{@{}c@{}}\textbf{LW (MHz)}\end{tabular} & \begin{tabular}[c]{@{}c@{}}\textbf{GS16-AWGN (dB)}\end{tabular} & \begin{tabular}[c]{@{}c@{}}\textbf{16QAM (dB)}\end{tabular} & \textbf{Fig.}  \\ 
\midrule
1 & 0.1 & 0.2 & \ref{fig:gmi_grid}(d) \\ 
2 & 0.35 & 0.2 & \ref{fig:gmi_grid}(e) \\ 
3 & 0.4 & 0.6 & \ref{fig:gmi_grid}(f) \\
\bottomrule
\end{tabular}
\label{tab:sg16}
\end{table}
\begin{table}[!h]
\renewcommand\arraystretch{1.5}%
\centering
\caption{32-point format optimisation\tablefootnote{Shown as the approximated shaping gains of GS32-RPN constellations with respect to the corresponding GS32-AWGN and 32QAM formats.}}
\begin{tabular}{cccc} 
\toprule
\begin{tabular}[c]{@{}c@{}}\textbf{LW (MHz)}\end{tabular} & \begin{tabular}[c]{@{}c@{}}\textbf{GS32-AWGN (dB)}\end{tabular} & \begin{tabular}[c]{@{}c@{}}\textbf{32QAM (dB)}\end{tabular} & \textbf{Fig.}  \\ 
\midrule
1 & $\approx{}0$ & 0.3 & \ref{fig:gmi_grid}(g) \\ 
2 & 0.4 & 0.6 & \ref{fig:gmi_grid}(h) \\ 
3 & 0.6 & 1 & \ref{fig:gmi_grid}(i) \\
\bottomrule
\end{tabular}
\label{tab:sg32}
\end{table}
\vspace{-12pt}
\newpage
\begin{table}[!h]
\renewcommand\arraystretch{1.5}%
\centering
\caption{64-point format optimisation\tablefootnote{Shown as the approximated shaping gains of GS64-RPN constellations with respect to the corresponding GS64-AWGN and 64QAM formats.}}
\begin{tabular}{cccc} 
\toprule
\begin{tabular}[c]{@{}c@{}}\textbf{LW (MHz)}\end{tabular} & \begin{tabular}[c]{@{}c@{}}\textbf{GS64-AWGN (dB)}\end{tabular} & \begin{tabular}[c]{@{}c@{}}\textbf{64QAM (dB)}\end{tabular} & \textbf{Fig.}  \\ 
\midrule
1 & 0.2 & 0.6 & \ref{fig:gmi_grid}(j) \\ 
2 & 0.5 & 1.0 & \ref{fig:gmi_grid}(k) \\ 
3 & 1.4 & 1.7 & \ref{fig:gmi_grid}(l) \\
\bottomrule
\end{tabular}
\label{tab:sg64}
\end{table}

In general, it can be observed that the GMI gap between the GS$M$-RPN and the other analyzed constellations increases as a function of the LW (i.e. the amount of RPN left after the CPE). Additionally, for the formats with cardinality of 16 and above, at fixed laser LW, the shaping gains increase with the order $M$ of the constellation. In order to better highlight these gains, Fig.~\ref{fig:shaping_gain} shows the shaping gains of GS$M$-RPN and GS$M$-AWGN over the corresponding $M$QAM format at a fixed GMI rate of 0.96$\times m$. The choice of this particular GMI rate is in line with the code rate adopted in the 400ZR specification and it will be substantiated by the results presented in Sec.~\ref{sec:fec_results}. Moreover, Tables \ref{tab:sg8}-\ref{tab:sg64} show the particular shaping gains of GS$M$-RPN formats over the other presented constellations and link them to Fig.~\ref{fig:shaping_gain}, which individually depicts the GMI performances of all the investigated SNR-RPN combinations. 
}

\begin{figure}[hb]
        \centering
        \begin{minipage}{\columnwidth}
            \centering
            \scalebox{.8}{
            \includegraphics[width=\textwidth]{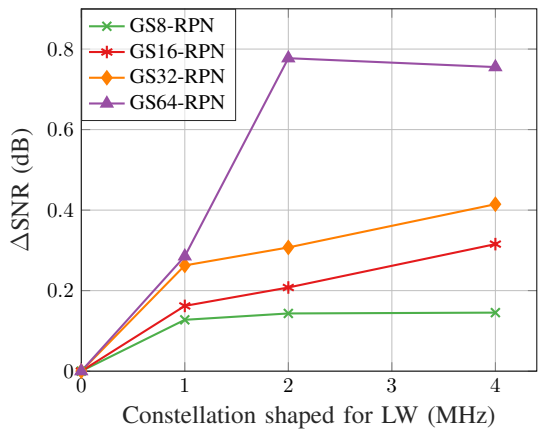}}
        \end{minipage}%
        \caption{The SNR penalty of GS$M$-RPN constellations in AWGN-only channel (i.e. GS$M$-RPN vs GS$M$-AWGN) at GMI=0.96$\times{}$m~bit/symbol at $60$~Gbaud.}
        \label{fig:curves_awgn}
        \vspace{-12pt}
\end{figure}
\subsection{Performance of the GS$M$-RPN in AWGN-only channel}
To provide a comprehensive overview of the presented GS technique, the GS$M$-RPN constellations were further verified in the AWGN-only channel. As shown in Fig. \ref{fig:curves_awgn}, the performance penalties of the GS$M$-RPN formats increase with the level of the RPN used for shaping. Compared to the GS$M$-AWGN formats, the penalties range from around $0.15$~dB for the 8-\textit{ary} constellation shaped for the RPN amount corresponding to 1~MHz lasers, up to around $0.8$~dB for GS64-RPN for $4$~MHz combined laser LWs.

\section{SD Hamming FEC results}
\label{sec:fec_results}
\begin{figure*}[!t]
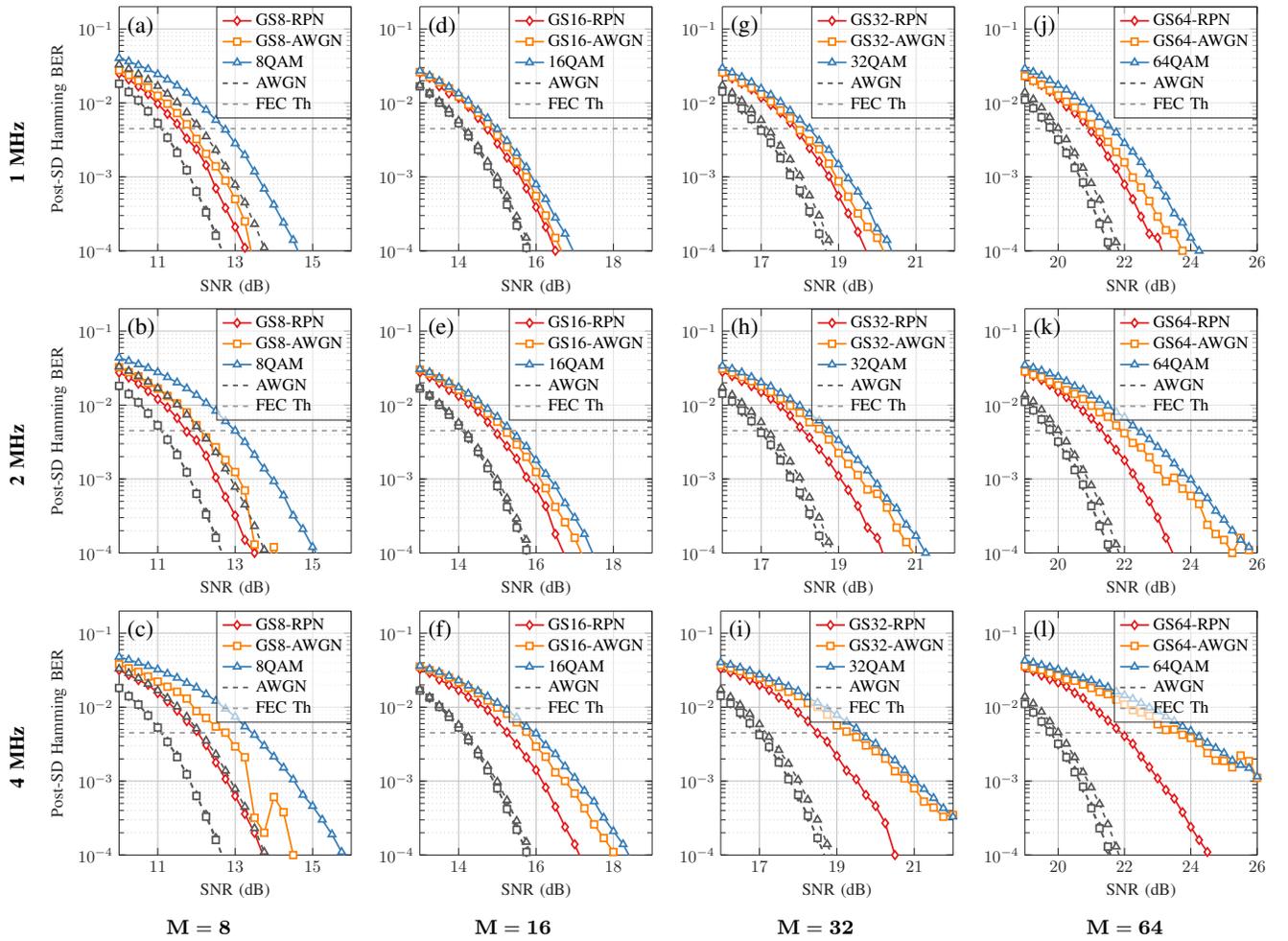

\centering

\begin{tabular}{c@{\hspace{.3cm}}c@{\hspace{.3cm}}c @{\hspace{.3cm}}c@{\hspace{.3cm}}c}
\rotatebox{90}{\makebox[1.6in][c]{\footnotesize{\textbf{1~MHz}}}} &
\includegraphics[height=1.6in]{8QAM/Curves/SD_Hamming_M=8_RPN=1.0e-04.tikz} &
\includegraphics[height=1.6in]{16QAM/Curves/SD_Hamming_M=16_RPN=1.0e-04.tikz} &
\includegraphics[height=1.6in]{32QAM/Curves/SD_Hamming_M=32_RPN=1.0e-04.tikz} &
\includegraphics[height=1.6in]{64QAM/Curves/SD_Hamming_M=64_RPN=1.0e-04.tikz} \\
\rotatebox{90}{\makebox[1.6in][c]{\footnotesize{\textbf{2~MHz}}}} &
\includegraphics[height=1.6in]{8QAM/Curves/SD_Hamming_M=8_RPN=2.1e-04.tikz} &
\includegraphics[height=1.6in]{16QAM/Curves/SD_Hamming_M=16_RPN=2.1e-04.tikz} &
\includegraphics[height=1.6in]{32QAM/Curves/SD_Hamming_M=32_RPN=2.1e-04.tikz} &
\includegraphics[height=1.6in]{64QAM/Curves/SD_Hamming_M=64_RPN=2.1e-04.tikz} \\
\rotatebox{90}{\makebox[1.6in][c]{\footnotesize{\textbf{4~MHz}}}} &
\includegraphics[height=1.6in]{8QAM/Curves/SD_Hamming_M=8_RPN=4.2e-04.tikz} &
\includegraphics[height=1.6in]{16QAM/Curves/SD_Hamming_M=16_RPN=4.2e-04.tikz} &
\includegraphics[height=1.6in]{32QAM/Curves/SD_Hamming_M=32_RPN=4.2e-04.tikz} &
\includegraphics[height=1.6in]{64QAM/Curves/SD_Hamming_M=64_RPN=4.2e-04.tikz} \\
&
\footnotesize{$\mathbf{M=8}$} &
\footnotesize{$\mathbf{M=16}$} &
\footnotesize{$\mathbf{M=32}$} &
\footnotesize{$\mathbf{M=64}$} \\
\end{tabular}

\caption{SD Hamming (128, 120) post-FEC performance for 8-\textit{ary} (a-c), 16-\textit{ary} (d-f), 32-\textit{ary} (g-i) and 64-\textit{ary} (j-l) constellations at 60~Gbaud. Modulation orders ($M$) are grouped in columns, whereas each row represents the combined laser linewidths of 1~MHz, 2~MHz and 4~MHz, respectively.}
\label{fig:grid_fec}
\end{figure*}
In order to validate the GMI-based optimization, the coded performance of the proposed GS constellations was investigated using a single-bit extended SD Hamming code (128,120) with rate \mbox{$\approx$0.94}, similar to the one recommended in \cite{oif}. In the 400ZR agreement the Hamming code is used as an inner SD code in combination with an outer HD staircase code \cite{Smith2012}. {\color{purple} The target output BER must be approximately equal to $4.5\cdot 10^{-3}$ to enable the outer staircase code (SCC) to achieve} a post-FEC BER below $10^{-15}$ (see \cite[Fig.~8]{Smith2012}). The SD-FEC decoder was implemented using a Chase3 SD algorithm \cite{Chase1972} which makes use of the channel bit-wise reliabilities (log-likelihood ratios). Unlike other more notable FEC schemes discussed in, e.g., \cite{Alvarado2016}, GMI is not necessarily a good indicator for the post-FEC performance of the SD Hamming code. However, at high rates, SNR gains at fixed GMI are still expected to translate well to post-Hamming FEC BER gains at $4.5\cdot 10^{-3}$. The latter gains are ultimately the relevant ones for a 400ZR-based system, as they also represent the overall post-FEC BER coding gains when a SCC at 6.7\% rate is used as an outer code. 

\begin{figure*}[!t]
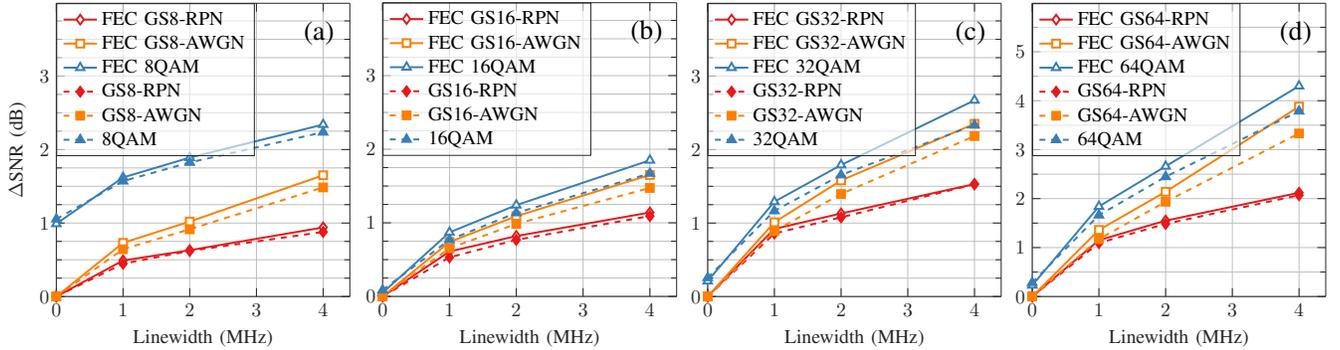

\centering
\begin{tabular}[!H]{c@{\hspace{.1cm}}c@{\hspace{.1cm}}c@{\hspace{.1cm}}c}
\includegraphics[height=1.832in]{8QAM/Curves/dSNR.tikz}
\includegraphics[height=1.835in]{16QAM/Curves/dSNR.tikz} &
\includegraphics[height=1.832in]{32QAM/Curves/dSNR.tikz} &
\includegraphics[height=1.832in]{64QAM/Curves/dSNR.tikz} \hfill\\
\end{tabular}
\caption{SNR penalty compared to to GS-AWGN constellation performance in an AWGN-only channel. The SNR gaps are measured at GMI= 0.96$\times{}$m~bit/symbol at $60$~Gbaud, for 8-\textit{ary} (a), 16-\textit{ary} (b), 32-\textit{ary} (c) and 64-\textit{ary} (d) constellations, respectively. The solid lines correspond to the GMI validation of each format, whereas the dashed lines are the SD Hamming (128,  120) post-FEC performance (described in Sec.~\ref{sec:fec_results}).}
\label{fig:coding_gain}
\vspace{-12pt}
\end{figure*}
{\color{purple}The post-Hamming BER estimation was performed using Monte-Carlo sampling of the channel} described in Sec.~\ref{sec:methodology}. {\color{purple} For each investigated modulation format and LW combination, a randomly generated sequence of $2^{20}$ symbols  was transmitted through the channel. It corresponds to  $2^{20}\cdot M$ samples of the equivalent bitwise channel, where $M$ is the constellation cardinality.} To obtain the required accuracy in the BER estimation, scrambling and interleaving of the available bit-wise channel samples were applied. This allowed us to emulate the transmission of the required number of codewords, potentially exceeding the number of Monte Carlo samples of the channel at low post-SD Hamming BER values. An ideal random interleaver with a decorrelation length larger than the codeword length was adopted to decorrelate bit errors within each received symbol. 

{\color{purple}The coded performance of the previously discussed GS constellations was examined as described in \ref{sec:methodology}-B.} In Fig.~\ref{fig:grid_fec}(a-c), the post-SD Hamming BER is shown for 8-\textit{ary} constellations as a function of the SNR for a single-channel 60~GBaud transmission in both AWGN channel and (a) 1 MHz, (b) 2 MHz, and (c) 4 MHz RPN channel. The compared constellations are GS8-AWGN, GS8-RPN, and 8QAM, as shown in Tab.~\ref{tab:sg8}. For the AWGN channel (black lines) both GS8-AWGN, and GS8-RPN have very similar performance, whilst 8QAM shows over 1 dB penalty. In the RPN channel, GS8-RPN outperforms all the other constellations for all investigated LWs, showing an increasing shaping gain compared to both GS8 and 8QAM as the LW increases. The SNR penalty $\Delta$SNR compared to AWGN performance for the 3 LWs examined is shown in Fig.~\ref{fig:coding_gain}(a). It can be noted that for the RPN channel the SNR penalty increases linearly (in dB) as a function of the LW. For the optimized GS8-RPN the $\Delta$SNR stays below 1 dB even for a 4 MHz LW, whereas for GS8-AWGN, and 8QAM $\Delta$SNR increases up to 1.6 dB and 2.4 dB, respectively, for a 4 MHz LW. Consistently with the shaping gains results shown in Fig.~\ref{fig:shaping_gain}, SNR penalties at a GMI rate of 0.96$\times m$ are also shown (dashed lines in Fig.\ref{fig:coding_gain}(a)).
It can be observed that there is good agreement between the SNR penalties computed at $4.5\cdot 10^{-3}$ post-SD Hamming BER and the ones computed at GMI=0.96$\times m$ bit/sym.

In Fig.~\ref{fig:grid_fec}(d-f), post-SD Hamming BER is shown for 16-\textit{ary} constellations as a function of the SNR for a single-channel 60 GBaud transmission in both AWGN channel and (a) 1 MHz, (b) 2 MHz, and (c) 4 MHz RPN channel. In this case, the different constellations investigated (GS16-AWGN, GS16-RPN, and 16QAM) show similar post-SD Hamming BER performance in the AWGN channel (black lines). However, as for the 8-\textit{ary} results the GS16-RPN constellation outperforms all other modulation formats for each investigated LW.
The SNR penalty performance of the $M=16$ formats is highlighted in Fig.~\ref{fig:coding_gain}(b). It can be seen that the SNR penalty compared to the performance of GS16 in AWGN increases up to 1.8 dB for 16QAM at 4 MHz LW, which is 0.6 dB lower than its 8-\textit{ary} counterpart. Moreover, the SNR gap between the different constellations is in general reduced compared to the 8-\textit{ary} case. In particular, at 1 MHz LW such a gap is 0.3 dB (GS16-RPN vs 16QAM) as opposed to 1.2 dB for the 8-\textit{ary} case. The $\Delta$SNR at 4 MHz LW for GS16-RPN is however increased to 1.2 dB (as opposed to 0.8 dB for GS8-RPN)

The results for the 32-\textit{ary} and 64-\textit{ary} constellations shown in Fig.~\ref{fig:grid_fec}(g-i) and \ref{fig:grid_fec}(j-l) are qualitatively in line with the 16-\textit{ary} case previously discussed. In particular, these results confirm that both the overall SNR penalty and SNR gap between modulation formats increase as a function of the constellation cardinality. It is worth noting that, as observed in Fig.~\ref{fig:coding_gain}(a-d), the GS$M$-RPN formats offer only a relatively marginal SNR improvement (close $\Delta$ SNR penalties values) compared to the corresponding GS$M$-AWGN format at 1 MHz LW. This improvement is approximately constant across all modulation formats and equal to $\approx$0.3 dB. On the other hand, at 4 MHz LW  $\Delta$SNR penalty incurred by 64QAM is 4.1 dB whereas it is only slightly above 2 dB for GS64-RPN, thus highlighting an SNR improvement for GS64-RPN of over 2 dB.

Interestingly, we observe that across all modulation formats and LWs the $\Delta$SNR measured in terms of post-SD Hamming FEC at 4.5$\cdot 10^{-3}$ is consistent with the ones measured at the selected GMI rate of 0.96$\times m$ bit/sym. We also note that this GMI threshold was observed to fit best (for the range GMI$\geq$ 0.9$\times m$ bit/sym) the post-SD Hamming SNR gains across all modulation formats and channels investigated. Although further investigation is needed, our results provide a first validation of the GMI-based shaping optimisation for the coded modulation scheme and the RPN channel studied in this work.

\section{Conclusion}
\label{sec:conclusion}
In this work, we introduced a new set of 2D geometrically shaped constellations that are simultaneously robust to both residual phase noise and AWGN. We observe that, in high SNR regimes, the 8, 16, 32 and 64-\textit{ary} GS$M$-RPN constellations outperformed both their GS$M$-AWGN and $M$QAM counterparts. Moreover, the GMI shaping results were consistent with the post-FEC verification in all of the presented cases. 

Furthermore, the presented gains increased proportionally to the cardinality of the constellation, and the variance of the residual phase noise after the pilot-aided CPE. Therefore, the most significant gain of $\approx$1.4~dB (between the GS$M$-RPN and GS$M$-AWGN formats), was recorded for 64-\textit{ary} constellations at $60$~Gbaud and a $4$~MHz combined laser LW. Moreover, when compared to 64QAM in the same channel, GS64-RPN has an improved SNR tolerance of approximately $1.7$~dB.

To conclude, we anticipate that this constellation shaping technique could be used either as a tool to relax LW and/or CPE requirements in low complexity coherent transmission systems or, by the same token, to improve the SNR tolerance in the presence of RPN. Additionally, by putting the presented results in the context of 400ZR, we have provided a low-complexity method to account for both the AWGN and RPN impairments in a comparable system. We also believe that the presented GS strategy can contribute as a potential road map for the short-reach optical fibre communication beyond 400~Gbit/s per channel. In the future, we plan to experimentally verify the performance of the presented formats and investigate the application of the presented theoretical gains in a practical system configuration.   

\section*{Acknowledgment}
The  authors  wish  to  extend  their  thanks  to  Drs  Christos Gkantsidis and Fotini Karinou from Microsoft Research for enlightening  discussions,  particularly  on  the  topic of the 400ZR implementation agreement.

\ifCLASSOPTIONcaptionsoff
  \newpage
\fi

\end{document}